\documentstyle[preprint,aps,prl,epsf,floats]{revtex}

\begin{document}

\draft

\tightenlines

\title{ 
     Fermi liquid theory 
     for the Anderson model out of equilibrium
}

\author{Akira Oguri}
\address{
          Department of Mathematics, Imperial College,  
         180 Queen's Gate, London SW7 2BZ, UK 
         \\
         and
         \\
         Faculty of Science, 
          Osaka City University, 
          Sumiyoshi-ku, Osaka 558-8585,
          Japan 
}

\date{December 21, 2000}

\maketitle

\begin{abstract}
The low-energy properties    
of the Anderson impurity 
are studied 
under a finite bias voltage $V$
using the perturbation theory in $U$ of Yamada and Yosida 
in the nonequilibrium Keldysh diagrammatic formalism.
The self-energy  
is calculated  exactly up to terms of order $\omega^2$, $T^2$ and $V^2$ 
using Ward identities.
The coefficients are defined with respect to the equilibrium ground state,
and contain all contributions of the perturbation series. 
From these results,
the nonlinear response of the current 
through the impurity has been deduced up to order $V^3$. 
\end{abstract}


\narrowtext
The Kondo effect \cite{Hewson}   
in quantum dots is a very active field of current research,
and the results of recent experiments \cite{Experiments} 
have been shown to be  in qualitative 
agreement with theoretical predictions \cite{Theories}. 
There is a new feature, the Kondo effect under a finite bias voltage $V$, 
which is specific to quantum dots  
and has no analogue in dilute magnetic alloys. 
So far, a number of theoretical methods have been used to
investigate out of equilibrium Kondo physics: 
 perturbation theory in the Coulomb 
interaction $U$ of the Anderson model\cite{HDW2}, 
the noncrossing approximation in the limit of $U\to \infty$\cite{WM}, 
the bosonization approach based on 
the Kondo model \cite{Schiller}, 
and so on. 
Although these works give some important insights
into the electron correlation in a driven system, 
further information is needed to clarify 
these new aspects of Kondo physics.

In the equilibrium state $V=0$ and linear response for $V \neq 0$, 
the low temperature properties of the Kondo physics 
can be described by a local Fermi liquid theory 
\cite{Nozieres,YamadaYosida}.  
The microscopic basis for this theory has been provided  
by a diagrammatic analysis 
based on the assumption 
that the ground state evolves continuously 
with the adiabatic switching-on of the Coulomb interaction $U$ 
\cite{YamadaYosida,Yoshimori}.
Some of the features found from the first few terms 
of the perturbation series  have been shown to hold 
to all orders in $U$, 
and can be expressed in terms of Fermi-liquid parameters 
which characterize the quasi-particle excitations.
The perturbation theory results have been confirmed 
by the exact Bethe ansatz solution \cite{ZlaticHorvatic}.
The local Fermi liquid theory also gives a description 
of the linear response of the current through 
small interacting systems \cite{AO}.
In the nonlinear response,
Hershfield {\em et al} have given the asymptotic form of the self-energy 
of the Anderson model to the second-order in $U$  
for the electron-hole symmetric case \cite{HDW2}. 
In this paper, we show that 
some of the features of the nonequilibrium state 
hold to all orders in $U$ without 
the assumption of the electron-hole symmetry. 
Our proof uses Ward identities for the derivative  
of the self-energy with respect to $V$, 
derived within a diagrammatic analysis based on  
the Keldysh formalism.
Consequently, the low-energy behavior of the 
self-energy $\Sigma^r(\omega)$ can be expressed 
in terms of Fermi-liquid parameters 
 for the equilibrium ground state;  
 Eqs.\ (\ref{eq:self_imaginary})-(\ref{eq:self_real}).  
The nonlinear response of the current $J$ has been calculated  
up to order $V^3$.
In the electron-hole symmetric case,
the expressions for $\Sigma^r(\omega)$ and $J$ can be written in 
simplified forms  
Eqs.\ (\ref{eq:self_symmetric_case})-(\ref{eq:current_symmetric_case}).

We start with the Anderson impurity connected to two reservoirs 
at the left ($L$) and right ($R$);
\begin{eqnarray}
H  
&=& 
 -\sum_{\lambda=L,R} 
 \sum_{\scriptstyle ij\in \lambda \atop \scriptstyle \sigma} 
         t_{ij}^{\lambda} \,  
         c^{\dagger}_{i \sigma} c^{\phantom{\dagger}}_{j \sigma}
 +  
   E_0 \sum_{\sigma} n_{0\sigma}
  +    U\, n_{0\uparrow}  n_{0\downarrow} 
\nonumber
\\
& &
-  \sum_{\sigma} v_L  \left[\,  
             c^{\dagger}_{0 \sigma}  c^{\phantom{\dagger}}_{-1 \sigma}
      +   c^{\dagger}_{-1 \sigma}  c^{\phantom{\dagger}}_{0 \sigma}
             \,\right] 
-  \sum_{\sigma}  v_R  \left[\,  
             c^{\dagger}_{1 \sigma} c^{\phantom{\dagger}}_{0 \sigma}
         +    c^{\dagger}_{0 \sigma}  c^{\phantom{\dagger}}_{1 \sigma}
             \,\right] 
.  
\label{eq:H}
\end{eqnarray}
Here 
$t_{ij}^{L}$  ($t_{ij}^{R}$) is the hopping matrix element
in the left (right) reservoir, 
$c^{\dagger}_{i \sigma}$ is creation operator of  
an electron with spin $\sigma$ at site $i$, 
 and
$n_{0\sigma} = c^{\dagger}_{0 \sigma} c^{\phantom{\dagger}}_{0 \sigma}$.
The couplings between the impurity and reservoirs are described  
by $v_L$ and $v_R$. 
The site indices are assigned to be 
 $i \leq -1$ for $i \in L$, and  $i \geq 1$ for $i \in R$.
The potential due to the bias voltage $V$ is 
included in the diagonal element $t_{ii}^{\lambda}$. 
We assume that $E_0$ is a constant independent of $V$, 
and take the chemical potential in the  
equilibrium state,  $\mu$, to be the origin of the energy.

The Keldysh formalism is described by the four types of Green's functions 
at the impurity site 
 \cite{KeldyshCaroliLandau}:
$G^{--} (t) \equiv -i \,
\langle \mbox{T}\,
c^{\phantom{\dagger}}_{0 \sigma}(t)\, c^{\dagger}_{0 \sigma}(0)
\rangle $, 
$G^{-+} (t) \equiv   i\, 
\langle 
c^{\dagger}_{0 \sigma}(0) \, c^{\phantom{\dagger}}_{0 \sigma}(t) 
\rangle$, 
$G^{+-} (t) \equiv  - i \,
\langle 
c^{\phantom{\dagger}}_{0 \sigma}(t) \, 
c^{\dagger}_{0 \sigma}(0)  
\rangle$,
$ G^{++}(t)\equiv -i \,
\langle \widetilde{\mbox{T}}\,
c^{\phantom{\dagger}}_{0 \sigma}(t)\, c^{\dagger}_{0 \sigma}(0)
\rangle$, 
and $c^{\phantom{\dagger}}_{0 \sigma}(t)$ is the Heisenberg operator 
which at $t=0$ coincides with the interaction representation. 
Furthermore, 
the retarded and advanced functions are given by 
$G^r \equiv G^{--}-G^{-+}$ and $G^a \equiv G^{--}-G^{+-}$. 
The average $\langle \cdots \rangle$ is taken 
with the density matrix at $t=0$. 
Initially at $t=-\infty$ 
the reservoirs are separated from the impurity 
and described by the thermal equilibrium 
of the chemical potential $\mu_L$ and $\mu_R$ 
with $\mu_L - \mu_R ={\sl e}V$. 
Then  $v_L$, $v_R$ and $U$ are switched on adiabatically.
In the noninteracting case $U=0$, 
the Fourier transform of the retarded and lesser Green's functions 
are given by 
\begin{eqnarray}
 G_{0}^{r}(\omega) &=&  
 \left[\,\omega-E_0 + i\, (\Gamma_L + \Gamma_R) \,\right]^{-1} 
,
\label{eq:G_0^r}
\\
G_{0}^{-+}(\omega) &=&  -\,
{ f_L(\omega)\, \Gamma_L + f_R(\omega)\, \Gamma_R 
  \over 
 \Gamma_L +\Gamma_R }
    \,
    \left[\,G_0^r(\omega)- G_0^a(\omega)\,\right] . 
\label{eq:G_0^-+}
\end{eqnarray}
Here $f_{L,R}(\omega) 
= f(\omega - \mu_{L,R})$ 
and
$f(\omega)=[\,e^{\omega/T}+1\,]^{-1}$. 
$\Gamma_{L} = \pi \rho_{L}^{\phantom{0}} v_{L}^2$,  
$\Gamma_{R} = \pi \rho_{R}^{\phantom{0}} v_{R}^2$, and  
$\rho_L^{\phantom{0}}$ ($\rho_R^{\phantom{0}}$) is 
the local density of states at the interface  $i= -1$ ($+1$).
We assume $\rho_{L,R}^{\phantom{0}}$ is a constant 
and the bandwidth is infinity, 
so that $\Gamma_{L,R}$ is a constant independent of $\omega$. 
From Eqs.\ (\ref{eq:G_0^r})-(\ref{eq:G_0^-+}),
other noninteracting Green's functions can be 
obtained using the properties  
$G^{-+}+G^{+-}=G^{--}+G^{++}$,  
$G^{++}(\omega)= - \{G^{--}(\omega)\}^*$,    
 and  $G^a(\omega) = \{G^r(\omega)\}^*$.  
Note that the distribution function is 
introduced through the $f_L \Gamma_L + f_R \Gamma_R$ dependence 
 of Eq.\ (\ref{eq:G_0^-+}).
It is not a symmetric function of $\mu_L$ and $\mu_R$ 
if $\Gamma_L \neq \Gamma_R$, 
 and thus the local charge 
 at the impurity site is affected 
by the inversion of the bias voltage.
The full Green's functions are described by the Dyson equation 
$
\{\mbox{\boldmath $G$}(\omega)\}^{-1} 
    =     
 \{\mbox{\boldmath $G$}_{0}(\omega)\}^{-1} 
-           \mbox{\boldmath $\Sigma$}(\omega)\, 
$;
\begin{eqnarray}
 \mbox{\boldmath $G$}_0 &=& \left[ 
 \matrix { G^{--}_0 & G^{-+}_0   \cr
           G^{+-}_0 & G^{++}_0  \cr  }
                             \right]
, 
\quad
\mbox{\boldmath $\Sigma$} \ = \ \left[ 
 \matrix { \Sigma^{--} & \Sigma^{-+}   \cr
           \Sigma^{+-} & \Sigma^{++}  \cr  }
                             \right] 
.
\end{eqnarray}
Here $\mbox{\boldmath $\Sigma$}(\omega)$ is the self-energy 
due to $U$, which can be defined using the diagrammatic representation
of the perturbation series \cite{HDW2,KeldyshCaroliLandau}. 
Furthermore,
the retarded function is given by    
$\{G^r\}^{-1} = \{G_0^r\}^{-1} - \Sigma^r$ with  
 $\Sigma^r \equiv \Sigma^{--} + \Sigma^{-+}$, and
 the current flowing through the impurity can be written 
in terms of the spectral 
function $A(\omega)\equiv  - (1/\pi)\,{\rm Im}\, G^r(\omega)$ \cite{MW};
\begin{eqnarray}
J  
 &=&  {2 {\sl e} \over h} \, 
 \frac{4\, \Gamma_L \Gamma_R}{\Gamma_R + \Gamma_L} 
\int_{-\infty}^{\infty} 
  d\omega  
   \left[\, f_L(\omega) - f_R(\omega) \, \right] 
   \pi A(\omega)  
.
\label{eq:current}
\end{eqnarray}
Generally 
 $A(\omega)$ is not symmetric 
against the inversion of the bias voltage
due to the $f_L \Gamma_L + f_R \Gamma_R$ dependence of $G^{\nu\nu'}_0$, 
and thus the nonlinear part of $J$ transforms 
to $-J$ only for $\Gamma_L = \Gamma_R$. 
In order to specify the deviations of $\mu_L$ and $\mu_R$ from 
the equilibrium value, 
we introduce the parameters $\alpha_L$ and $\alpha_R\,$:  
$\mu_L \equiv \alpha_L \,{\sl e}V$ and
$\mu_R \equiv - \alpha_R \,{\sl e}V$ with $\alpha_L + \alpha_R=1$.
The inversion of the bias voltage is described by 
the transformation $(V, \alpha_L, \alpha_R) 
\Rightarrow (-V, \alpha_R, \alpha_L)$.  
In the following, we will discuss the behavior  
of $\Sigma^r(\omega)$ in the vicinity of the equilibrium 
up to order $V^2$ by evaluating the first 
and second derivatives with respect to $V$.

We now consider the first derivative. 
At $V=0$,  
Eq.\ (\ref{eq:G_0^-+}) 
is written as $G_{0eq}^{-+}(\omega) \equiv
\left.G_{0}^{-+}(\omega)\right|_{V=0} = -f(\omega)\,
[\,G_{0eq}^r(\omega)- G_{0eq}^a(\omega)\,]$ with  
 $G_{0eq}^{r}(\omega) \equiv G_{0}^{r}(\omega)$.
The derivative of $G_{0}^{\nu \nu'}$ with 
respect to $V$ can be written  
in terms of the Green's functions for the equilibrium state;  
\begin{eqnarray}
\left.
{\partial G_{0}^{\nu \nu'}(\omega) 
\over
\partial ({\sl e}V)}\right|_{V=0}
&=&
 -\, \alpha 
\left(
{\partial \over \partial \omega}
+ {\partial \over \partial E_0}
\right)
G_{0eq}^{\nu \nu'}(\omega)  ,
\label{eq:derivative_1b}
\\ 
\left(
{\partial \over \partial \omega}
+ {\partial \over \partial E_0}
\right)
G_{0eq}^{\nu \nu'}(\omega) 
 &=&
 - {\partial f(\omega) \over \partial \omega}
 \left[\, 
G_{0eq}^{r}(\omega) -
G_{0eq}^{a}(\omega) 
\,\right] .
\label{eq:derivative_1a}
\end{eqnarray}
Here  $ \alpha \equiv (\alpha_L \Gamma_L - \alpha_R \Gamma_R)/ 
 (\Gamma_L+ \Gamma_R) $ and
 $G_{0eq}^{\nu \nu'}(\omega) \equiv 
\left. G_{0}^{\nu \nu'}(\omega)\right|_{V=0}$ 
with $\nu, \nu' = +, -$. 
Note that $\alpha$ becomes zero 
for   
$\alpha_L = \Gamma_R/(\Gamma_L+\Gamma_R)$ 
and 
$\alpha_R = \Gamma_L/(\Gamma_L+\Gamma_R)$. 
The derivative of $\mbox{\boldmath $\Sigma$}(\omega)$ can be carried out 
using the diagrammatic analysis, i.e.,  
taking the derivative of $G_{0}^{\nu \nu'}$'s
within the diagrams 
for $\mbox{\boldmath $\Sigma$}(\omega)$ \cite{Yoshimori}.
In the closed loops of the diagrams,
the frequency of $G_{0}^{\nu \nu'}(\omega_i)$  can be shifted 
to $\omega_i + \omega$ without changing the value of 
$\mbox{\boldmath $\Sigma$}(\omega)$ because the frequencies of 
the closed loops are integrated out. 
Therefore, it is possible to assign the frequencies of the diagrams  
so that all the noninteracting Green's functions 
possess the external frequency $\omega$ in the argument.  
Then, the derivative $\partial /\partial({\sl e}V)$ can be 
replaced by $-\alpha\,(\partial/\partial \omega + \partial/\partial E_0)$ for 
all the internal $G_{0}^{\nu \nu'}(\omega_i+\omega)$'s using 
Eq.\ (\ref{eq:derivative_1b}), and thus 
\begin{equation}
\left.
{\partial  
\mbox{\boldmath $\Sigma$}(\omega) 
\over
\partial ({\sl e}V)}\right|_{V=0} 
 =  
 -\,
 \alpha  
\left(
{\partial \over \partial \omega}
+ {\partial \over \partial E_0}
\right)
\mbox{\boldmath $\Sigma$}_{eq}(\omega) 
.
\label{eq:derivative_self_1}
\end{equation}
Here $\mbox{\boldmath $\Sigma$}_{eq}(\omega) 
\equiv \left. \mbox{\boldmath $\Sigma$}(\omega) \right|_{V=0}$. 
We will assign the label \lq$eq$' to the subscript of the functions 
for the equilibrium state.
Note that Eq.\ (\ref{eq:derivative_self_1}) 
holds at any temperatures since no assumption has been made 
for the value of $T$ so far. 
The same relation holds for $\Sigma^r(\omega)$,
and operating $\partial/ \partial \omega$ to 
the expression corresponding to Eq.\ (\ref{eq:derivative_self_1}),
we obtain
 \begin{equation}
 \left.
 {\partial^2  
 \Sigma^r(\omega) 
 \over
 \partial \omega \partial ({\sl e}V)}\right|_{V=0} 
 =  
   -\,   \alpha \,
{\partial \over \partial \omega}
\left(
{\partial \over \partial \omega}
+ {\partial \over \partial E_0}
\right)
\Sigma_{eq}^r(\omega) 
.
\label{eq:derivative_self_1.5}
\end{equation}
In the equilibrium state,  
the imaginary part of the self-energy behaves as 
$\mbox{Im}\,\Sigma_{eq}^r(\omega) \propto [\omega^2 + (\pi T)^2]$ 
for small $\omega$ and $T$ \cite{YamadaYosida,Yoshimori,Note}. 
Thus, substituting this into the right-hand-side of 
Eq.\ (\ref{eq:derivative_self_1.5}), 
we find that $\mbox{Im}\,\Sigma^r(\omega)$ has 
the cross-term $\omega{\sl e}V$.
Similarly, differentiating  Eq.\ (\ref{eq:derivative_self_1}) 
with respect to $T$, we find that
there is no cross-term 
of $T{\sl e}V$ in the imaginary part. 
At $T=0$ and $V=0$, the usual diagram technique in terms of  
 $G^{--}_{eq}$ is applicable,  
and the causal element of Eq.\ (\ref{eq:derivative_self_1}) can be 
written in terms of the vertex part using Eq.\ (\ref{eq:derivative_1a}) 
 \cite{Yoshimori}, 
\begin{eqnarray}
\left( {\partial \over \partial \omega} 
+ {\partial \over \partial E_0} \right)
 \Sigma_{eq}^{--}(\omega) 
&=& \sum_{\sigma'}\int {d\omega' \over 2\pi i}\, 
\Gamma_{\sigma\sigma';\sigma'\sigma}(\omega,\omega';\omega',\omega)
\,\delta G(\omega') 
\left\{-{\partial f(\omega') \over \partial \omega'} \right\}
\nonumber \\
&=&  
{\delta G(0) \over 2\pi i}\, 
\sum_{\sigma'}
\Gamma_{\sigma\sigma';\sigma'\sigma}(\omega,0;0,\omega) \;.
\label{eq:Ward_1}
\end{eqnarray}
Here 
$\Gamma_{\sigma\sigma';\sigma'\sigma}(\omega,\omega';\omega',\omega)$ 
is the antisymmetrized total vertex part of the $T=0$ formalism, 
$\delta G(\omega) 
\equiv G_{eq}^{r}(\omega) - G_{eq}^{a}(\omega)$,  
and $-\partial f(\omega) /\partial \omega = \delta(\omega)$. 
We note that at $T=0$ the differential coefficients  
are related to the enhancement factors of the physical quantities: 
 $\widetilde{\gamma}  
 \equiv 1 - 
 {\partial \Sigma_{eq}^r(\omega) / \partial \omega}|_{\omega=0} $,
$\widetilde{\chi}_{c}  \equiv
\widetilde{\chi}_{\uparrow\uparrow} 
+ \widetilde{\chi}_{\uparrow\downarrow} 
=
 1 + {\partial \Sigma_{eq}^r(0) / \partial E_0}
$,
and 
$\widetilde{\chi}_{s}  \equiv
\widetilde{\chi}_{\uparrow\uparrow} 
- \widetilde{\chi}_{\uparrow\downarrow} 
$,
where 
$\widetilde{\chi}_{\sigma\sigma'} 
= \delta_{\sigma\sigma'}  -  
\left.
{\partial \Sigma_{eq,\sigma}^r(0) / \partial h_{\sigma'} }
\right|_{h_{\sigma'}=0}
$
with $h_{\sigma'}$ being an external field described by 
$H_{ex} 
= -  \Sigma_{\sigma} h_{\sigma}n_{0\sigma}
$.
Furthermore, there are some exact relations among 
these parameters \cite{YamadaYosida,Yoshimori}: 
$\widetilde{\chi}_{\uparrow\uparrow} =  \widetilde{\gamma}$ 
and 
 $\widetilde{\chi}_{\uparrow\downarrow} = 
 -A_{eq}(0)\,\Gamma_{\uparrow\downarrow; \downarrow\uparrow}(0,0;0,0)$.

Next we consider the second derivative with respect to $V$ in 
the similar way. 
The second derivative of 
the noninteracting Green's functions 
can be written as 
\begin{eqnarray}
\left.
{\partial^2 G_{0}^{\nu \nu' }(\omega) 
\over
\partial ({\sl e}V)^{2}}\right|_{V=0}
 &=&
\kappa 
\left(
{\partial \over \partial \omega}
+ {\partial \over \partial E_0}
\right)^{2}
G_{0eq}^{\nu \nu'}(\omega)
, 
\label{eq:derivative_2b}
\\
\left(
{\partial \over \partial \omega}
+ {\partial \over \partial E_0}
\right)^{2}
G_{0eq}^{\nu \nu'}(\omega) 
 &=& -\,
 {\partial^2 f(\omega) \over \partial \omega^2}
 \left[\, 
G_{0eq}^{r}(\omega) -
G_{0eq}^{a}(\omega) 
\,\right]
.
\label{eq:derivative_2a}
\end{eqnarray}
Here 
$
\kappa \equiv 
   {
  (\alpha_L^2 \Gamma_L + \alpha_R^2 \Gamma_R)/ 
 (\Gamma_L+ \Gamma_R})  
$.
The second derivative of $\mbox{\boldmath $\Sigma$}(\omega)$ can be  
 evaluated by operating 
$\left\{\partial /\partial({\sl e}V) \right\}^2$ to 
$G_{0}^{\nu \nu'}(\omega_i + \omega)$'s within   
the diagrams for $\mbox{\boldmath $\Sigma$}(\omega)$. 
The contributions are classified into two groups 
according to  
whether $\left\{\partial /\partial({\sl e}V) \right\}^2$
operates to, ({\em i\,}) two different electron lines,
 or ({\em ii\,}) a single electron line.
In the class ({\em i\,}), 
each of the first derivatives $\partial /\partial({\sl e}V)$ 
can be replaced by 
$ -\alpha\,(\partial/\partial \omega + \partial/\partial E_0)$. 
In the class ({\em ii\,}),
the second derivative  $\partial^2 /\partial({\sl e}V)^2$ can be replaced by
$\kappa\,(\partial/\partial \omega + \partial/\partial E_0)^2$.  
Therefore,
\begin{eqnarray}
\left.
{\partial^2  
\mbox{\boldmath $\Sigma$}(\omega) 
\over
\partial ({\sl e}V)^2}\right|_{V=0} 
&=& 
 \alpha^2 \,
  (\, \mbox{\boldmath $C$}_I  + \mbox{\boldmath $C$}_{II} \,)
 \, + \,  (\, \kappa -  \alpha^2  \,)\,  \mbox{\boldmath $C$}_{II}
\nonumber
\\
&=& 
\alpha^2
\left(
{\partial \over \partial \omega}
 + {\partial \over \partial E_0}
\right)^2 
\mbox{\boldmath $\Sigma$}_{eq}(\omega) 
 +    
{ \Gamma_L\,\Gamma_R 
  \over \left( \Gamma_L+ \Gamma_R \right)^2}
   \  \widehat{D}^2 
\mbox{\boldmath $\Sigma$}_{eq}(\omega) 
.
\label{eq:derivative_self_2_with_CII}
\end{eqnarray}
Here 
$\alpha^2 \,\mbox{\boldmath $C$}_I$ and 
$\kappa\,\mbox{\boldmath $C$}_{II}$ denote 
the contributions of the classes ({\em i\,}) and ({\em ii\,}), respectively. 
Note that  
$\mbox{\boldmath $C$}_I + \mbox{\boldmath $C$}_{II}  
=\left(
{\partial / \partial \omega}
+ {\partial / \partial E_0}
\right)^2 \mbox{\boldmath $\Sigma$}_{eq}(\omega)$. 
$
\mbox{\boldmath $C$}_{II}=
\widehat{D}^2 
\mbox{\boldmath $\Sigma$}_{eq}(\omega) 
$, and 
$\widehat{D}^2$ denotes the functional operation 
carrying out $(\partial /\partial \omega + \partial/ \partial E_0 )^2$ 
for all the single $G_{0}^{\nu \nu'}$.
At $T=0$, the usual zero-temperature formalism is applicable 
for the causal element $\widehat{D}^2 \Sigma_{eq}^{--}(\omega)$. 
The functional operation to pick up 
one electron line from the diagrams for $\Sigma_{eq}^{--}(\omega)$ has been 
used to derive Eq.\ (\ref{eq:Ward_1}) \cite{Yoshimori}.
Carrying out the same operation and using 
Eq.\ (\ref{eq:derivative_2a}) 
instead of Eq.\ (\ref{eq:derivative_1a}),
we obtain 
\begin{eqnarray}
   \widehat{D}^2  \Sigma_{eq}^{--}(\omega) 
&=& 
\sum_{\sigma'}\int {d\omega' \over 2\pi i}\, 
\Gamma_{\sigma\sigma';\sigma'\sigma}(\omega,\omega';\omega',\omega)
\,\delta G(\omega') 
\left\{-{\partial^2 f(\omega') \over \partial \omega'^2} \right\}
\nonumber \\
&=&  
-{\delta G(0) \over 2\pi i}\, 
\sum_{\sigma'} \left. {\partial \over \partial \omega'}
\Gamma_{\sigma\sigma';\sigma'\sigma}(\omega,\omega';\omega',\omega)
\right|_{\omega'=0} 
 -  
{\delta G'(0) \over 2\pi i}
\sum_{\sigma'}
\Gamma_{\sigma\sigma';\sigma'\sigma}(\omega,0;0,\omega)
.
\label{eq:Ward_2x} 
\end{eqnarray}
Here $\delta G'(\omega) = \partial \delta G(\omega)/\partial \omega$,
and 
$-\partial^2 f(\omega) /\partial \omega^2 = \delta'(\omega)$. 
Note that 
$
\Gamma_{\sigma\sigma';\sigma'\sigma}(\omega,\omega';\omega',\omega)
 = 
\Gamma_{\sigma'\sigma;\sigma\sigma'}(\omega',\omega;\omega,\omega')
$ due to the antisymmetric property of the total vertex.
Furthermore, owing to the rotational symmetry of the spins, 
the first term of Eq.\ (\ref{eq:Ward_2x})
can also be written as
$\sum_{\sigma'}
(\partial / \partial \omega')\,
\Gamma_{\sigma\sigma';\sigma'\sigma}(\omega',\omega;\omega,\omega')
$.
Now we consider the limit $\omega \to 0$. 
As shown by \'{E}liashberg quit generally,
the vertex part 
$\Gamma_{\sigma\sigma';\sigma'\sigma}
(\omega,\omega';\omega',\omega)$ has 
singularities at $\omega -\omega'=0$ and 
$\omega+\omega'=0$ 
\cite{Eliashberg}.
In Fig.\ \ref{fig:vertex}, 
the diagrams contribute to the singularities 
for small $\omega$ and $\omega'$ are shown:   
(a)-(b) and (c)
yield the $|\omega-\omega'|$ and $|\omega+\omega'|$ dependence, 
respectively.
In the figures, 
the solid line denotes the full Green's function, and 
the square denotes the vertex part  
with zero frequencies  
$\Gamma_{\uparrow\downarrow;\downarrow\uparrow}(0,0;0,0)$. 
Note that  $\Gamma_{\uparrow\uparrow;\uparrow\uparrow}(0,0;0,0)=0$ because of 
the Pauli principle. 
Therefore, 
the derivative of the vertex part has a discontinuity 
of the form
$ ({\partial / \partial \omega'})\, |\omega - \omega'|
 =  -\, \mbox{sgn}(\omega -\omega') $, 
and the limit $\omega,\, \omega' \to 0$ depends on 
which frequency is first taken to be zero. 
These discontinuities appear in the imaginary part,
and for small $\omega$ and $\omega'$  
the contributions of the diagrams (a)-(c) 
are written as 
\begin{eqnarray}
& &   
\sum_{\sigma'}  
{\partial \over \partial \omega'}
\, \mbox{Im}\, 
\Gamma_{\sigma\sigma';\sigma'\sigma}(\omega,\omega';\omega',\omega)
\nonumber \\
&=&    
-\,
\left|\Gamma_{\uparrow\downarrow;\downarrow\uparrow}(0,0;0,0)
\right|^2 
\nonumber \\
& & 
\times \, \mbox{Im} \left[ \,
2 \int {d \omega'' \over 2\pi i}\,
G_{eq}^{--}(\omega'') \,
{\partial  \over \partial \omega'} G_{eq}^{--}(\omega -\omega'+ \omega'')
+ \int {d \omega'' \over 2\pi i}\,
G_{eq}^{--}(\omega'') \,
{\partial  \over \partial \omega'} G_{eq}^{--}(\omega +\omega'-\omega'')
\,\right]
\nonumber
\\
&=&    
-\,
\pi\,\widetilde{\chi}_{\uparrow\downarrow}^2 \,
\left[\, 
- 2\,\mbox{sgn}(\omega'-\omega) \,
+ \mbox{sgn}(\omega' +\omega) \,
\,\right].
\label{eq:Ward_2_Im_diagram}
\end{eqnarray}
Note that 
$G_{eq}^{--}(\omega) = 
 \mbox{Re}\,G_{eq}^{r}(\omega) 
+i\, \mbox{Im}\,G_{eq}^{r}(\omega)\, \mbox{sgn}(\omega)
$ 
at $T=0$. 
Thus, taking the limit $\omega'\to 0$ first,  
we obtain 
\begin{eqnarray}
\lim_{\omega \to 0}\,
\mbox{Im}\, \widehat{D}^2  \Sigma_{eq}^{--}(\omega) 
=  
- 3\,\pi A_{eq}(0)\, \widetilde{\chi}_{\uparrow\downarrow}^2
\, \mbox{sgn}(\omega). 
\label{eq:Ward_2_Im}
\end{eqnarray}
The opposite limit of Eq.\ (\ref{eq:Ward_2_Im_diagram}) 
corresponds to the $\omega^2$ term of 
$\mbox{Im}\,\Sigma_{eq}^{--}(\omega)$, 
which is obtained by operating $\partial/\partial \omega$ 
to Eq.\ (\ref{eq:Ward_1}) as  
$
\left.
({\partial^2  
/ \partial \omega^2})\, 
\mbox{Im}\,
\Sigma_{eq}^{--}(\omega)
 \right|_{\omega \to 0} =
- \pi A_{eq}(0)\, \widetilde{\chi}_{\uparrow\downarrow}^2
\, \mbox{sgn}(\omega)
$ \cite{YamadaYosida,Yoshimori}.
On the other hand, 
the real part of the vertex part does not show these discontinuities,  
and the limit is independent of the ways for $\omega,\,\omega' \to 0$. 
Thus, 
in the real part of Eq.\ (\ref{eq:Ward_2x}), 
we can take the limit $\omega \to 0$ first. 
Then using Eq.\ (\ref{eq:Ward_1}), we obtain
\begin{eqnarray}
 \mbox{Re}\,  \widehat{D}^2 
 \Sigma_{eq}^{--}(0) 
&=&
- \left.
{\partial \over \partial \omega}
\left( 
{\partial 
\over \partial \omega} 
+ {\partial \over \partial E_0} \right)
\mbox{Re}\,
 \Sigma_{eq}^{--}(\omega)
 \right|_{\omega= 0}  
  -  \, {A_{eq}'(0) \over A_{eq}(0)}\,
   \widetilde{\chi}_{\uparrow \downarrow} .
\label{eq:Ward_2_Re}
\end{eqnarray}
Consequently,
using Eqs.\ (\ref{eq:derivative_self_2_with_CII}) and 
(\ref{eq:Ward_2_Im})-(\ref{eq:Ward_2_Re}),
$\left.\partial^2 \Sigma_{eq}^{--}/\partial({\sl e}V)^2\right|_{V=0}$ 
can be expressed in terms of the enhancement factors  
for the equilibrium ground state.

    We now summarize the results to 
show the low-energy behavior of $\Sigma^r(\omega)$.
Since the imaginary part has no $T{\sl e}V$ dependence as mentioned,
the result, which is valid up to $\omega^2$, $V^2$, and $T^2$, 
can be written as
\begin{eqnarray}
\mbox{Im}\, \Sigma^r(\omega) =   
      -\, { \pi A_{eq}(0)\, 
       \widetilde{\chi}_{\uparrow\downarrow}^2\over 2}\,
        \left[\,
            \left(\,\omega - 
              \alpha\, {\sl e}V\, 
              \right)^2 
              +  { 3\,\Gamma_L \Gamma_R 
                 \over \left( \Gamma_L + \Gamma_R \right)^2}
                \,({\sl e}V)^2 
              +(\pi T)^2  
           \,\right] 
 + \cdots 
 . 
\label{eq:self_imaginary}
\end{eqnarray}
Note that electron-hole symmetry has not been assumed so far.
The general expression of the real part is rather long. 
At $T=0$ it takes the form 
\begin{eqnarray}
\mbox{Re}\, \Sigma^r(\omega)  &=&  
        \Sigma_{eq}^r(0) 
     \, + \, (1-\widetilde{\chi}_{\uparrow\uparrow})\, \omega
     \, - \,  
            \alpha\, 
        \widetilde{\chi}_{\uparrow\downarrow}\,
             {\sl e}V 
 \, + \, {b\over 2}\,\omega^2
             \, - \,
            \alpha  
              \left( b  -   
              {\partial\, \widetilde{\chi}_{\uparrow\uparrow}
             \over \partial E_0}             \right)
            \omega \,{\sl e}V 
     \nonumber \\
& & \, +  \, 
     {1\over 2}\, 
         \left[\, 
                  \alpha^2  
                \left( b    
                   -   
                  {\partial\, \widetilde{\chi}_{s}
                  \over \partial E_0}\, 
                 \right)
\,- \, 
          {\Gamma_L \Gamma_R 
                 \over \left( \Gamma_L+ \Gamma_R \right)^2}
           \left(b  -  {\partial\, \widetilde{\chi}_{\uparrow\uparrow}
                              \over \partial E_0} 
                        +  {A_{eq}'(0) \over A_{eq}(0)} 
                     \,\widetilde{\chi}_{\uparrow\downarrow}\,
                \right)
           \,\right]
                 ({\sl e}V)^2 
 + \cdots 
                 .
\label{eq:self_real}
\end{eqnarray}
Here
$b \equiv  \left. (\partial^2 / \partial \omega^2)\,
      \mbox{Re}\,\Sigma_{eq}^r(\omega)
\right|_{\omega=0}$.
In the electron-hole symmetric case  
($E_0 = -U/2$, $\Gamma_L = \Gamma_R$, and $\alpha_L=\alpha_R=1/2$), 
the real part Eq.\ (\ref{eq:self_real}) simplifies 
 and we obtain the expression,
\begin{eqnarray}
\Sigma^r(\omega) &=&
      (1-\widetilde{\chi}_{\uparrow\uparrow})\, \omega
      \, - \, 
      i\, {\widetilde{\chi}_{\uparrow\downarrow}^2 \over 2\Delta}
        \,\left[\,\omega^2 + {3 \over 4}\,({\sl e}V)^2 +(\pi T)^2
       \,\right]
        + \cdots 
 ,
\label{eq:self_symmetric_case}
\end{eqnarray}
which for $V=0$ corresponds  
to the result of Yamada and Yosida \cite{YamadaYosida}.
Here $\Delta = \Gamma_R + \Gamma_L$.
We note that in the symmetric case  the Bethe ansatz solution for 
$\widetilde{\chi}_{\sigma \sigma'}$ 
can be expanded in power series of $U$  
which converges absolutely for any finite $U$ 
\cite{ZlaticHorvatic}, and 
the second-order result 
of $\mbox{Im}\, \Sigma^r(\omega)$ of Hershfield {\em et al\/} \cite{HDW2} 
is reproduced through the first term of the perturbation series 
$\widetilde{\chi}_{\uparrow\downarrow} = -U/(\pi\Delta) + \cdots$
\cite{YamadaYosida,ZlaticHorvatic}. 
In the general formula 
Eqs.\ (\ref{eq:self_imaginary})-(\ref{eq:self_real}),
the contributions of the higher order terms are described through 
the parameters $\widetilde{\chi}_{\sigma\sigma'}$, 
$A_{eq}(0)$,  $\Sigma_{eq}^r(0)$, and $b$.

Using the results of the self-energy 
Eqs.\ (\ref{eq:self_imaginary})-(\ref{eq:self_real}), 
the low energy behavior 
of  $A(\omega)$ is obtained correctly 
up to terms of order $\omega^2$ and $V^2$.
Then, using Eq.\ (\ref{eq:current}), 
the current is evaluated exactly up to order $V^3$ 
as $J=g_1 V+g_2 V^2+g_3 V^3+\cdots$. 
Note that  
the current conservation is fulfilled up to order $V^3$ 
because all contributions of the perturbation series 
are taken into account for the coefficients $g_1$, $g_2$ and $g_3$.
Generally, the explicit form of $g_3$ is rather long 
reflecting the lengthy expression of 
the real part Eq.\ (\ref{eq:self_real}).
Specifically, in the electron-hole symmetric case, $g_2$ vanishes, 
and $g_1$ and $g_3$ can be expressed in terms of 
the two parameters  
 $\widetilde{\chi}_{\uparrow\uparrow}$ and
 $\widetilde{\chi}_{\uparrow\downarrow}$, and the resulting expression of
 the current is 
\begin{eqnarray}
J 
 &=&  {2 {\sl e}^2 \over h} \, V 
  \left[\, 1 \,
  - 
                \, {\widetilde{\chi}_{\uparrow\uparrow}^2
                  + 2\, \widetilde{\chi}_{\uparrow\downarrow}^2 
                  \over 3} 
                 \left( {\pi\,T \over \Delta}\right)^2
             - 
                \, {\widetilde{\chi}_{\uparrow\uparrow}^2
                  + 5\, \widetilde{\chi}_{\uparrow\downarrow}^2 
                  \over 12} 
                 \left( {{\sl e}V \over \Delta}\right)^2 
 + \cdots 
 \,\right] .             
\label{eq:current_symmetric_case}
\end{eqnarray}
Furthermore, 
in the Kondo limit ($U \to \infty$ with $E_0 = -U/2$), 
the charge fluctuation is suppressed $\widetilde{\chi}_c \to 0$ 
and the low-energy behavior of the current is characterized by
the single parameter $\widetilde{\chi}_s$ as 
\begin{equation}
J  
 =  { 2 {\sl e}^2  \over h} \,  V 
  \left[\, 1 \, - \,
                {\widetilde{\chi}_{s}^2 \over 4}
                 \left( {\pi\,T \over \Delta}\right)^2
   - \, {\widetilde{\chi}_{s}^2 \over 8}  
   \left( {{\sl e}V \over \Delta} \right)^2 
  + \cdots \, \right] .              
\label{eq:current_Kondo}
\end{equation}
In this limit the spin susceptibility is determined by 
the Kondo temperature as $\widetilde{\chi}_s \propto \Delta /T_K$, 
and thus the $V^2$ and $T^2$ dependence of the 
differential conductance $dJ/dV$ is scaled by $T_K$.

In conclusion, we have derived 
the Ward identities for the derivative of the self-energy with respect 
to $V$ based on the perturbation theory in $U$.
Using the identities,
we have calculated exactly the low-energy behavior of 
the self-energy and nonlinear response of the current 
in the nonequilibrium steady state.


I would like to thank A. C. Hewson for useful comments and discussions,
 and H. Ishii and  S. Nonoyama for valuable discussions. 
I wish to thank the Newton Institute in Cambridge
for hospitality during my stay 
on the programme  \lq Strongly Correlated Electron Systems'.
This work is supported by the Grant-in-Aid 
for Scientific Research from the Ministry of Education, 
Science and Culture, Japan.

\end{document}